# Self-Organized Growth of Nanoparticles on a Surface Patterned by a Buried Dislocation Network

F. Leroy[1*], G. Renaud[1] A. Letoublon[1], R. Lazzari[2], C. Mottet[3], J. Goniakowski[2]

[1]*CEA-Grenoble, Département de Recherche Fondamentale sur la Matière Condensée SP2M/NRS, 17, rue des Martyrs, F-38054 Grenoble, France.*

[2]*Institut des Nanosciences de Paris, CNRS UMR 7588-Universités Paris VI-VII, 140, rue de Lourmel, 75015 Paris, France*

[3]*CRMCN-CNRS, Campus de Luminy, case 913, 13288 Marseille Cedex 9, France*

Abstract

The self-organized growth of Co nanoparticles with 10 nm periodicity was achieved at room temperature on a Ag(001) surface patterned by an underlying dislocation network, as shown by real time, *in situ* Grazing Incidence Small and Wide Angle X-ray Scattering. The misfit dislocation network, buried at the interface between a 5nm-thick Ag thin film and a MgO(001) substrate, induces a periodic strain field on top of the surface. Nucleation and growth of Co on tensile areas are found as the most favorable sites as highlighted by Molecular Dynamic simulations.

**PACS numbers: 81.16.Rf, 61.10.Eq, 68.55.Ac, 61.72.Hh**

The fabrication of ordered metal and semiconductor nanoparticles on solid surfaces with uniform and controllable size and shape and with a high spatial density is an important challenge as it may find applications in future nanoelectronics [1], , ultra-high density recording [2] and nanocatalysis [3]. Two different routes have been taken towards nanopatterning: one by developing new scanning techniques with nanometer resolution [4], the other by transferring the periodicity of spontaneous self-organised surface patterns, such as surface reconstructions [5], to nanoparticles superlattices. The latter approach avoids the broadening of the size distribution inherent to the random processes of deposition and diffusion on flat substrates, and offers an economic and parallel way to realise high density integration, for which lithography techniques find their limits. The self-organized growth (SOG) of magnetic materials is an appealing technique for instance in view of magnetic recording.

It has been shown that strain-relief patterned substrates such as dislocation networks (DN) can serve as templates for growing uniform regular-spaced nanostructures [6-8]. Due to long-range repulsive interactions, the dislocations arrange into highly ordered periodic patterns. The inhomogeneous strain field experienced by diffusing atoms gives rise to heterogeneous nucleation and growth at specific sites, yielding a well-ordered nanoparticles superlattice [9,10]. However, up to now, organized clusters were realized on very specific closed packed (111) surface of fcc crystals for which ordering could only be obtained at low deposition temperature. Moreover the nanostructures were only of one or two atomic layers high [6], while producing structures with larger thickness is mandatory to increase blocking temperature up to room temperature (RT).

In this letter we present a new method based on (001) surfaces nanostructured by a buried misfit edge dislocation network whose periodic strain field emerges at the surface. We show that the trapping energy of atoms is large enough to allow an ordering of nanostructures at RT. The cobalt/silver interface is chosen because it is a test bed for magnetic nanoparticles, as Co exhibits a three-dimensional (3D) growth on Ag(001) and because it does not alloy [11] at RT. To modulate the surface strain field, a Ag film was grown on a MgO(001) substrate. Due to the cube on cube epitaxial relationship and the 3% lattice mismatch between Ag and MgO(001), strain relaxation occurs *via* a square misfit dislocation network with a period of $D\sim 10$ nm [12].

If scanning tunneling microscopy (STM) is the technique of choice to study SOG, it suffers in the present case from its inability to probe the DN internal characteristics. In this context, depth sensitivity as well as statistical information on a macroscopic scale to characterize the order quality are crucial to understand SOG processes. For that reason, we resorted to *in situ* Grazing Incidence Small Angle X-ray Scattering (GISAXS) [13], which has become a new tool to investigate nanometer scale order close to a surface.

The experiments were performed at the European Synchrotron Radiation Facility (ESRF), on the BM32 beamline [14], delivering a monochromatic (0.06888nm) X-ray beam, and using a newly developed set-up allowing to perform GISAXS, Grazing Incidence X-ray Diffraction (GIXD) and X-ray Reflectivity (XR) measurements on the same sample, *in situ,* in Ultra High Vacuum (UHV), at different growth stages. The incident X-ray beam impinges on the surface at a grazing angle $\alpha_i$ and the scattered intensity is recorded as a function of out-of-plane angle $\alpha_f$ and in-plane angle $\theta_f$ (Fig. 2a). These angles allow defining the reciprocal space coordinates $Q_\perp$ and $Q_{//}$, respectively perpendicular and parallel to the surface. The small angle scattering was collected on a one megapixel 16-bit X-ray CCD camera located at ~1.7 m downstream the sample. The MgO(001) single crystal was prepared following a procedure yielding a high quality surface [15]. Ag and Co were respectively deposited, under UHV (base pressure $5.10^{-11} mbar$), using a Knudsen cell and an electron bombardment source while performing GISAXS, GIXD and XR measurements. The deposition rates were *in situ* calibrated with a quartz microbalance and XR. A 100 nm-thick 2D Ag film was first grown on MgO(001) at RT and then annealed at 900K, yielding a Ag(001) film of high crystalline quality (mosaic spread smaller than 0.05°), and exhibiting a well ordered interfacial DN as revealed by many diffraction satellites around the Ag Bragg peaks [12]. Then the film was thinned *in situ* by ion bombardment until ~5 nm thickness (as determined by XR), while the thinning process was monitored by Ag(110) anti-Bragg GIXD measurements (Fig. 1). The temperature during bombardment was then precisely determined to stay in a regime of step retraction [16], thus keeping large terraces (100 nm) and low roughness.

Before proceeding to the Co growth, a detailed (nano-) crystallographic study of the substrate by GISAXS was performed. Similarly to GIXD experiments, this first step is necessary to analyze after deposition the interferences between the waves scattered by the substrate and those scattered by the Co nanostructures. Figures 2(a) and (b) display two GISAXS images measured on the Ag/MgO(001) film with the incident X-ray beam respectively parallel to the $\langle 110 \rangle$ and $\langle 100 \rangle$ MgO(001) crystalline axes. Sharp scattering rods in the $Q_{//}$ direction reveal a periodic nanopattern of four-fold symmetry. The large terrace size demonstrates that the GISAXS signal does not arise from a surface superstructure, but rather from the buried DN. The in plane rod positions correspond, as expected [12], to dislocation lines oriented along the <110> substrate directions, with a period D=10.95 nm. The quality of the dislocation network is revealed by the correlation length of the superlattice, $D \times (Q_{//} / \Delta Q_{//}) = 170$ nm, as deduced from the Full Width Half Maximum ($\Delta Q_{//}$) of the scattering rods. A quantitative analysis of the $(1/\Lambda,1/\Lambda,L)$ and $(2/\Lambda,0,L)$ ($\Lambda \sim 37$ atoms) scattering rods was performed in the Distorted Wave Born Approximation (DWBA) framework [18] since it provides an internal description of the strain field, both parallel and perpendicular to the interface. As expected from the isotropic linear elasticity theory applied to a perfect misfit dislocation network [17], an exponential damping of the strain field as a function of the vertical $z$ coordinate, was assumed inside the film ($\theta_f$), and in the substrate ($\theta_s$).

$$\theta_f = \gamma_f \exp(-z/\lambda) \text{ and } \theta_s = \gamma_s \exp(z/\lambda), \tag{1}$$

where $z = 0$ at the interface and $\lambda$ is the attenuation length of the strain associated with the misfit dislocations. The elastic constants of both materials are taken into account through $\gamma_f$ and $\gamma_s$. For both $(1/\Lambda,1/\Lambda,L)$ and $(2/\Lambda,0,L)$ scattering rods (Fig. 2c), a value of $\lambda = 1.05$ nm was deduced as qualitatively expected [17].

GISAXS measurements were then performed during the growth of Co on this nanostructured template for different substrate temperatures and Co growth rate. Co was finally deposited at room temperature and at a very low rate ($4 \times 10^{-3}$ nm/min), respectively to decrease the thermal energy with respect to the DN nucleation trapping potential [5], and to increase the diffusion length of Co atoms and thus their probability to find a nucleation site. No ordering of Co clusters was found at higher temperatures or deposition rates. From the very beginning of the growth (0.04 nm), the subtracted GISAXS images (after and before Co deposition) display intensity oscillations along the DN scattering rods with a damped sinusoidal shape [19] (Fig.3a). The oscillation amplitude increases with deposition time, reaches a maximum for an equivalent Co deposited thickness of 0.19 nm, and then decreases (Fig.3b). The period, equals to 5 nm, is a signature of the height difference between the Co clusters and the interfacial DN. Most importantly, these oscillations reveal the SOG of Co clusters, since an interference effect can only occur if the phase shift between the waves scattered by the Co clusters and those scattered by the DN is well defined, *i.e.* if the Co clusters are well localized with respect to the dislocations positions. Indeed, the intensity is the sum of three terms: the intensity scattered by the DN, the one scattered by the Co islands, and the interference term between both:

$$I = |F_{DN}|^2 + |F_{Co}|^2 + 2 \times F_{DN} F_{Co} \cos\left(\vec{Q_{//}} \vec{d_{//}} + Q_\perp d_\perp\right) \qquad (2)$$

where $F_{DN}$ (resp. $F_{Co}$) is the form factor of the DN (respectively of the Co clusters), and $\vec{d_{//}}$ and $d_\perp$ are the parallel and perpendicular coordinates of the Co clusters with respect to the dislocation intersection lines (Fig. 3c).

As Co clusters are very small ($|F_{Co}| << |F_{DN}|$), the $|F_{Co}|^2$ term can be neglected in equation (2). Thus, the interference term, which contains the information on the Co clusters location, is simply obtained by the subtraction of GISAXS measurements after and before Co deposition. On the basis of the strain field symmetries, two high-symmetry sites are possible for the Co clusters: above the dislocation lines crossing or in-between. In order to discriminate between the two possible locations, the interference effect along the $(1/Λ,1/Λ,L)$ rod was simulated in the DWBA framework. The Co clusters were taken as cylinders whose height is an integer number of Co atomic layers (AL). The best fit was unambiguously obtained for the clusters located above the dislocation core (Fig. 3b), and a height of 2 AL.

To interpret this result, theoretical calculations of the adsorption energy of a Co atom on a nanostructured thin film of Ag(001) on MgO(001) were performed. The energetic model is based on a semi-empirical tight binding potential for metal-metal interactions and on a potential fitted to *ab initio* calculations for the metal-MgO(001) ones (details are given in ref. [20]). The bare strained thin film of Ag(001) on MgO(001) was first investigated. Molecular dynamic simulations were performed on Ag atoms whereas the oxide surface was frozen. Assuming periodic boundary conditions, the simulations were made on a superlattice unit cell with a lateral periodicity of 9.82 nm ($Λ$~34 atoms) and a thickness of 5 nm (24 AL). The adsorption site of Ag was located on top of the O as demonstrated elsewhere [21]. Simulations clearly show that the Ag surface exhibits alternating tensile and compressive areas (Fig. 4), the former being located on top of the dislocations crossing. The Co adsorption on top of the tensile zones is found more favorable than on compressive ones with an energy difference of 60 meV, which increases up to 80 meV for thinner films (20 AL). Such a result is comparable to other calculations performed on Pt(111) strained surfaces [22, 23]. We show here that taking into account a realistic strained surface (the silver film nanostructured by the buried DN on MgO(100) substrate), the energy difference between the adsorption sites is much higher than the thermal activation energy available at room temperature. If we suppose that the first atom adsorption is a good indicator for further cluster nucleation [22], such an energy difference should contribute to the cluster organization. Moreover, from a kinetic point of view, the diffusion barrier by hopping is found to be significantly lower on compressive sites as compared to tensile ones (by 60 meV) which reinforces the tendency for Co clusters to nucleate on tensile sites [23].

The position of Co clusters being determined, we resorted to a detailed analysis of the amplitude of the interference term as function of deposition time. From equation (2), the amplitude is proportional to $|F_{Co}|$. Assuming a cylindrical shape and a radius $R$ proportional to the square root of time to account for the linear increase of volume during deposition, the amplitude evolution is in good agreement with the one expected from the measured deposition rate (Fig. 3b and 3c). The linear increase of the amplitude at the beginning of the growth is a consequence of the small island size: the damping of the Co form factor with momentum transfer can be neglected and the resulting amplitude is proportional to the island volume *i.e.* to the deposition time. The decreasing amplitude at the end of the deposition is due to Co clusters getting closer to coalescence for which the intensity should decrease to zero.

To conclude, we have shown that the spatially dependent surface strain induced by a misfit dislocation network buried as far as 5 nm below a Ag(001) surface allows controlling the growth of Co clusters at RT, leading to self-organized growth. This result is supported by Molecular Dynamic Simulations. We believe that this method could be used for many different systems, metal thin films being favored with respect to semi-conductor ones because of the dislocation mobility necessary to reach the equilibrium state.

Acknowledgements

We would like to acknowledge the invaluable help of Marion Ducruet and Ameline Crémona for the many sample preparations and of Tobias Schülli during some measurements, as well as the ESRF and BM32 staff for beam availability.References

*New address: MPI-FKF, Heisenbergstr. 1, D-70569 Stuttgart. Germany. email: f.leroy@fkf.mpg.de

1 A.O. Orlov *et al.*, Science **277**, 928 (1997)

2 S. Sun, C. Murray, D. Weller, L. Folks and A. Moser, Science **287**, 1989 (2000)

3 M. Valden, X. Lai and D.W. Goodman, Science **281**, 1647 (1998)

4 D. Eiglet and E. Schweizer, Nature **344**, 524 (1990)

5 H. Brune, Surf. Sci. Rep. **31**, 121 (1998)

6 B. Voigtländer, G. Meyer, N. M. Amer, Phys. Rev. B **44**, 10354 (1991)

7 H. Brune, M. Giovanni, K. Bromann and K. Kern, Nature **394**, 451 (1998)

8 R.Q. Hwang and M.C. Bartelt, Chem. Rev. **97**, 1063 (1997)

9 B. Yang, F. Liu and M.G. Lagally, Phys. Rev. Lett. **92**, 025502 (2004)

10 A. Bourret, Surf. Sci. **432**, 37 (1999)

11 B. Degroote, J. Dekoster, G. Langouche, Surf. Sci. **452**, 172 (2000)

12 G. Renaud, P. Guénard and A. Barbier, Phys. Rev. B **58**, 7310 (1998)

13 G. Renaud et al., Science, **300**, 1416 (2003).

14 http://www.esrf.fr.

15 O. Robach, G. Renaud and A. Barbier, Surf. Sci. 410, 227 (1998).

16 C. Boragno et al., Phys. Rev. B. 65, 153406 (2002).

17 R. Bonnet and J.L. Verger-Gaugry, Philos. Mag. A **66**, 849 (1992).

18 S. K. Sinha, E. B. Sirota, S. Garoff and H.B. Stanley, Phys. Rev. B **38**, 2297 (1988).

19 The damping is actually due to the curvature of the Ewald's sphere.8

Figures caption

FIG. 1. Ag(110) anti-Bragg peak intensity versus time (logarithmic basis). (a) at 470K before ion bombardment (IB) the intensity is steady. (b): during IB at 470K, the intensity decreases and oscillates, exhibiting a layer by layer ablation process (see inset). (c) at 570K the intensity increases to reach a steady, maximum value.

FIG. 2. (a) Scattering geometry of GISAXS and experimental pattern with the incident beam along the MgO[110] direction. The intensity is represented on a logarithmic scale. The direct and reflected beams are hidden by a vertical beam-stop. The first and second order DN scattering rods are indicated by arrows. Inclined rods at 54° with respect to the surface normal arise from Ag(111) facets (present because the Ag films is not perfectly 2D, but made of very large flat Ag islands with a top (001) surface and small (111) facets at the edges, as seen by cross section TEM). (b) Same as (a), but with the incident beam along [100]. (c) Cuts along $Q_\perp$ of the scattering rods of (a) ($\Delta$) and (b) (O) (multiplied by 10 for clarity) and best fits (solid lines).

FIG. 3. (a) Experimental interference pattern (after subtraction of the substrate's one) with the incident beam along the [110] direction, for a 0.14 nm-thick Co deposition. The intensity is represented with a linear scale, the $Q_{//}$ (resp. $Q_\perp$) axis ranges from −1 to 1nm$^{-1}$ (resp. 0 to 3.26nm$^{-1}$). Oscillations along the rods are clearly visible. (b) Intensity of the interference term versus $Q_\perp$, for different deposition times (symbols) with best fits (see text). A vertical translation proportional to the time of deposition has been introduced for clarity. Last (top) curve: interference term (circle) and best fit (solid line) for 0.14nm-thick Co deposition, and a simulated interference term (dashed line) for Co clusters located at the center of the unit cell.

(c) Schematic representation of the Co clusters position with respect to the dislocation intersection lines (d) Oscillations amplitude versus time. Experimental data (□) and best fit (solid line) for the (1/Λ,1/Λ,L) DN rod as well as simulated curve of Co clusters radius versus time (short-dashed and ◊) for 2AL high clusters.

Fig. 4. Atomic stress map of the Ag nanostructured film in top view (top panel) and cut view in the (010) plane (bottom). Color code from red/dark to blue/light corresponds to compressive to tensile atomic sites. The stress amplitude being exponentially attenuated from the interface to the surface, the color code between the surface and the rest of the Ag film is decorrelated in order to distinguish the stress oscillations on the surface.

Figures

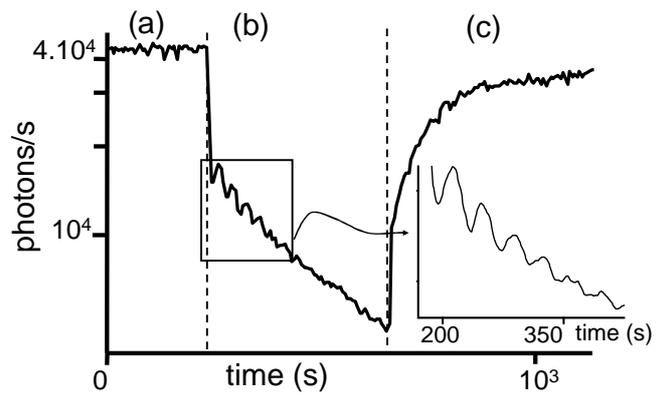

FIG1

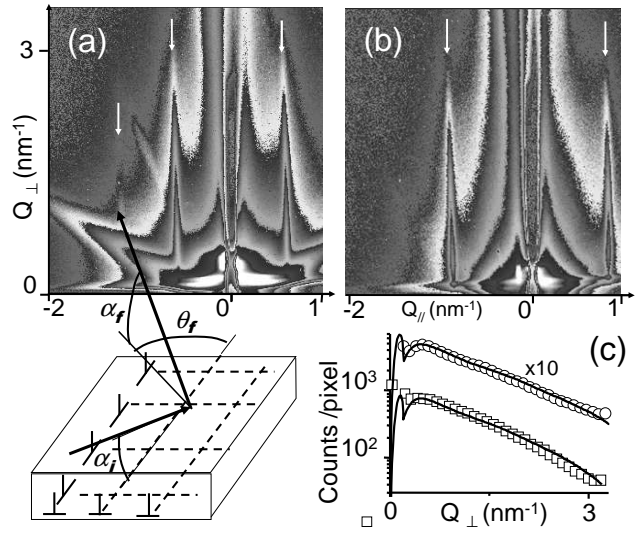

FIG 2

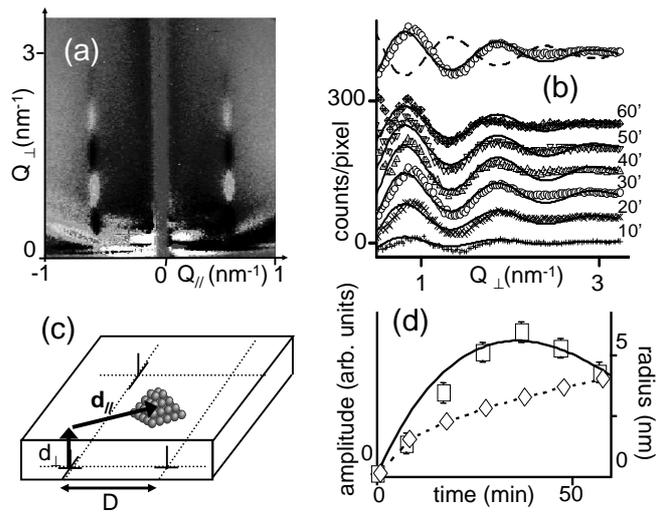

FIG 3

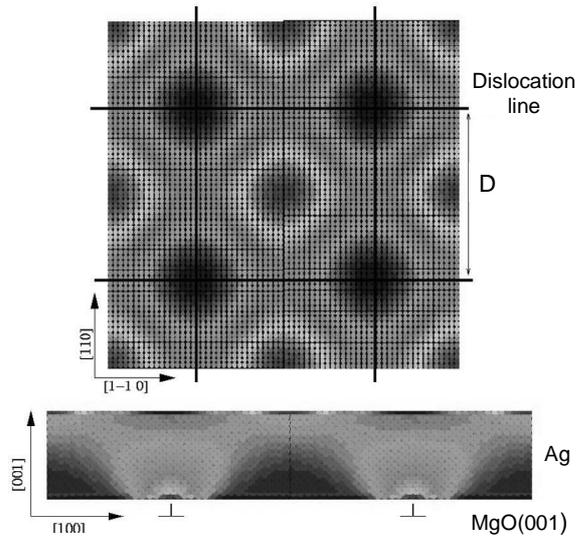

FIG 4